\documentclass[amsmath,amssymb,twocolumn,prc,showkeys,nofootinbib,floatfix,superscriptaddress]{revtex4-1}

\usepackage[T1]{fontenc}
\usepackage{amsmath}
\usepackage{amssymb}
\usepackage{mathtools}
\usepackage{booktabs}
\usepackage{siunitx}
\usepackage[referable]{threeparttablex}
\usepackage{longtable}
\usepackage{hyperref}
\usepackage{graphicx}
\usepackage{color}

\usepackage{breakcites}
\usepackage{float}
\usepackage{soul}
\usepackage{amsmath}
\usepackage{graphicx}
\usepackage{dcolumn}
\usepackage{bm}
\usepackage{natbib}
\usepackage{tikz}
\usepackage{color}
\usepackage{multirow}
\usepackage{subfigure}
\usepackage{longtable}
\usepackage{dcolumn}

\begin{document}

\title{Development and operation of an electrostatic time-of-flight detector for the Rare RI storage Ring}

\author{D.~Nagae}
\email[Corresponding author: ]{daisuke.nagae@riken.jp}
\affiliation{RIKEN Nishina Center, RIKEN, 2-1 Hirosawa, Wako-shi, Saitama 351-0198, Japan}
\author{Y.~Abe}
\affiliation{RIKEN Nishina Center, RIKEN, 2-1 Hirosawa, Wako-shi, Saitama 351-0198, Japan}
\author{S.~Okada}
\affiliation{Institute of Physics, University of Tsukuba, 1-1-1 Tennodai, Tsukuba-shi, Ibaraki 305-8571, Japan}
\author{S.~Omika}
\affiliation{Department of Physics, Saitama University, Shimo-Okubo 255, Sakura-ku Saitama-shi, Saitama, 338-8570 Japan}
\author{K.~Wakayama}
\affiliation{Department of Physics, Saitama University, Shimo-Okubo 255, Sakura-ku Saitama-shi, Saitama, 338-8570 Japan}
\author{S.~Hosoi}
\affiliation{Department of Physics, Saitama University, Shimo-Okubo 255, Sakura-ku Saitama-shi, Saitama, 338-8570 Japan}
\author{S.~Suzuki}
\affiliation{Institute of Physics, University of Tsukuba, 1-1-1 Tennodai, Tsukuba-shi, Ibaraki 305-8571, Japan}
\author{T.~Moriguchi}
\affiliation{Institute of Physics, University of Tsukuba, 1-1-1 Tennodai, Tsukuba-shi, Ibaraki 305-8571, Japan}
\author{M.~Amano}
\affiliation{Institute of Physics, University of Tsukuba, 1-1-1 Tennodai, Tsukuba-shi, Ibaraki 305-8571, Japan}
\author{D.~Kamioka}
\affiliation{Institute of Physics, University of Tsukuba, 1-1-1 Tennodai, Tsukuba-shi, Ibaraki 305-8571, Japan}
\author{Z.~Ge}
\affiliation{RIKEN Nishina Center, RIKEN, 2-1 Hirosawa, Wako-shi, Saitama 351-0198, Japan}
\affiliation{Department of Physics, Saitama University, Shimo-Okubo 255, Sakura-ku Saitama-shi, Saitama, 338-8570 Japan}
\author{S.~Naimi}
\affiliation{RIKEN Nishina Center, RIKEN, 2-1 Hirosawa, Wako-shi, Saitama 351-0198, Japan}
\author{F.~Suzaki}
\affiliation{RIKEN Nishina Center, RIKEN, 2-1 Hirosawa, Wako-shi, Saitama 351-0198, Japan}
\author{N.~Tadano}
\affiliation{Department of Physics, Saitama University, Shimo-Okubo 255, Sakura-ku Saitama-shi, Saitama, 338-8570 Japan}
\author{R.~Igosawa}
\affiliation{Department of Physics, Saitama University, Shimo-Okubo 255, Sakura-ku Saitama-shi, Saitama, 338-8570 Japan}
\author{K.~Inomata}
\affiliation{Department of Physics, Saitama University, Shimo-Okubo 255, Sakura-ku Saitama-shi, Saitama, 338-8570 Japan}
\author{H.~Arakawa}
\affiliation{Department of Physics, Saitama University, Shimo-Okubo 255, Sakura-ku Saitama-shi, Saitama, 338-8570 Japan}
\author{K.~Nishimuro}
\affiliation{Department of Physics, Saitama University, Shimo-Okubo 255, Sakura-ku Saitama-shi, Saitama, 338-8570 Japan}
\author{T.~Fujii}
\affiliation{Department of Physics, Saitama University, Shimo-Okubo 255, Sakura-ku Saitama-shi, Saitama, 338-8570 Japan}
\author{T.~Mitsui}
\affiliation{Department of Physics, Saitama University, Shimo-Okubo 255, Sakura-ku Saitama-shi, Saitama, 338-8570 Japan}
\author{Y.~Yanagisawa}
\affiliation{RIKEN Nishina Center, RIKEN, 2-1 Hirosawa, Wako-shi, Saitama 351-0198, Japan}
\author{H.~Baba}
\affiliation{RIKEN Nishina Center, RIKEN, 2-1 Hirosawa, Wako-shi, Saitama 351-0198, Japan}
\author{S.~Michimasa}
\affiliation{Center for Nuclear Study, The University of Tokyo, 2-1 Hirosawa, Wako, Saitama 351-0198, Japan}
\author{S.~Ota}
\affiliation{Center for Nuclear Study, The University of Tokyo, 2-1 Hirosawa, Wako, Saitama 351-0198, Japan}
\author{G.~Lorusso}
\affiliation{Department of Physics, University of Surrey, Guildford, Surrey GU2 7XH, United Kingdom}
\author{Yu.~A.~Litvinov}
\affiliation{GSI Helmholtzzentrum f\"{u}r Schwerionenforschung GmbH, 64291 Darmstadt, Germany}
\author{A.~Ozawa}
\affiliation{Institute of Physics, University of Tsukuba, 1-1-1 Tennodai, Tsukuba-shi, Ibaraki 305-8571, Japan}
\author{T.~Uesaka}
\affiliation{RIKEN Nishina Center, RIKEN, 2-1 Hirosawa, Wako-shi, Saitama 351-0198, Japan}
\author{T.~Yamaguchi}
\affiliation{Department of Physics, Saitama University, Shimo-Okubo 255, Sakura-ku Saitama-shi, Saitama, 338-8570 Japan}
\affiliation{Tomonaga Center for the History of the Universe, University of Tsukuba, 1-1-1 Tennodai, Tsukuba-shi, Ibaraki, 305-8571, Japan}
\author{Y.~Yamaguchi}
\affiliation{RIKEN Nishina Center, RIKEN, 2-1 Hirosawa, Wako-shi, Saitama 351-0198, Japan}
\author{M.~Wakasugi}
\affiliation{RIKEN Nishina Center, RIKEN, 2-1 Hirosawa, Wako-shi, Saitama 351-0198, Japan}

\date{\today}

\begin{abstract}
An electrostatic time-of-flight detector named E-MCP has been developed for 
quick diagnostics of circulating beam and timing measurement in mass spectrometry at the Rare-RI Ring in RIKEN.
The E-MCP detector consists of a conversion foil, potential grids, and a microchannel plate.
Secondary electrons are released from the surface of the foil when a heavy ion hits it.
The electrons are accelerated and deflected by 90$^\circ$ toward the microchannel plate by electrostatic potentials.
A thin carbon foil and a thin aluminum-coated mylar foil were used as conversion foils.
We obtained time resolutions of 69(1)~ps and 43(1)~ps (standard deviation) for a $^{84}$Kr beam 
at an energy of 170~MeV/u when using the carbon and the aluminum-coated mylar foils, respectively. 
A detection efficiency of approximately 90\% was obtained for both foils.
The E-MCP detector equipped with the carbon foil was installed inside the Rare-RI Ring 
to confirm particle circulation within a demonstration experiment on mass measurements of nuclei around $^{78}$Ge
produced by in-flight fission of uranium beam at the RI Beam Factory in RIKEN.
Periodic time signals from circulating ions were clearly observed.
Revolution times for $^{78}$Ge, $^{77}$Ga, and $^{76}$Zn were obtained.
The results confirmed successful circulation of the short-lived nuclei inside the Rare-RI Ring.
\end{abstract}


\maketitle
\sloppy


\section{Introduction}
Nuclear masses are important for understanding nuclear properties related to nuclear structure, 
such as shell closures and the onset of deformation,
weak and electromagnetic interactions, and borders of nuclear existence.
Systematic studies on nuclear masses are crucial 
for understanding the synthesis of heavy elements in astrophysical processes.
To measure the masses of nuclei located far from stability,
a storage ring called Rare-RI Ring~\cite{ozawa} was constructed at the RI Beam Factory in RIKEN~\cite{yano}.
It employs the isochronous mass spectrometry technique~\cite{haus1, haus2}, 
with the goal of determining masses with a precision of $10^{-6}$ in less than 1-ms measurement time.
The mass of a particle of interest is obtained by using the following equation:
\begin{equation}
\label{eq:m/q}
\frac{m_{1}}{q_{1}} = \frac{m_{0}}{q_{0}} \frac{T_{1}}{T_{0}} \sqrt{\frac{1-\beta_{1}^{2}}{1-(\frac{T_{1}}{T_{0}}\beta_{1})^{2}}}
\end{equation}
where $m_{0,1}/q_{0,1}$ denote the mass-to-charge ratios of the reference and the particle of interest, respectively; 
$T_{0,1}$ are their revolution times; 
$\beta_{1}$ is the velocity of the particle of interest. 
Since the isochronous condition is fulfilled for one nucleus only (such as the reference particle),
a correction of the revolution time by the velocity of the particle of interest is indispensable~\cite{YHZ, geis}.

\begin{figure}[htb]
\centering
\includegraphics[clip, width=8cm]{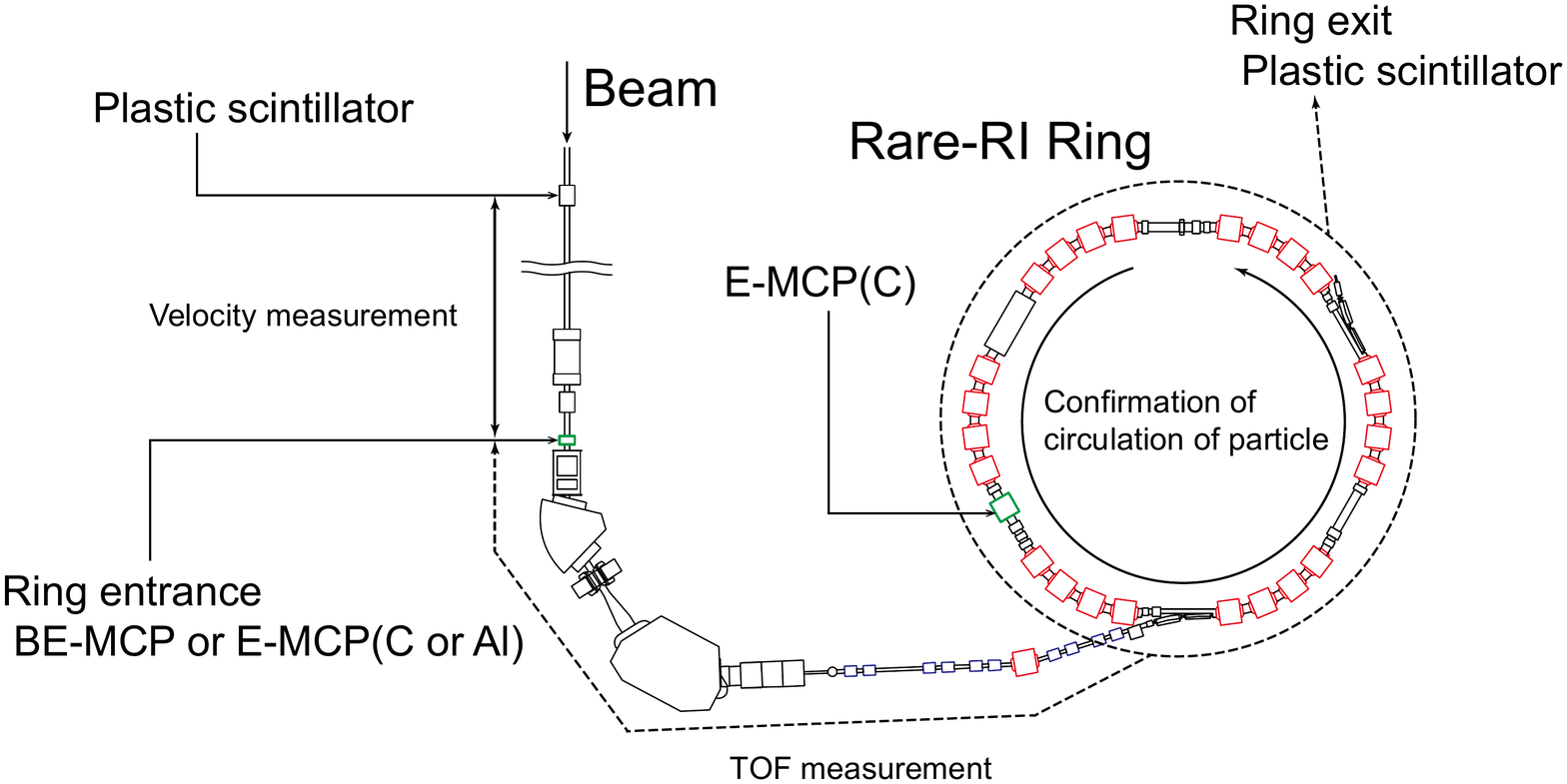}
\caption{Schematic view of the Rare-RI Ring including the location of the BE-MCP and the E-MCP detectors (see text). 
The velocity is measured upstream of the ring,
and the TOF is measured between the entrance and exit of the ring as indicated by the dashed line.
Either the BE-MCP or the E-MCP(C or Al) is installed at the entrance of the ring.
The E-MCP(C) was installed inside the ring.
Here, E-MCP(C) and E-MCP(Al) indicate the E-MCP detector equipped with carbon foil
and the aluminum-coated mylar foil, respectively.}
\label{fig:ring}
\end{figure}

Figure \ref{fig:ring} shows a schematic view of the Rare-RI Ring, the location of the time-of-flight (TOF) detectors, 
and the mass-measurement scheme.
The particles are injected into the ring one by one, stored for approximately $0.7$~ms,
and then extracted.
The mass is determined by measuring 
the TOF between the entrance and exit of the ring (instead of the revolution time, see Figure \ref{fig:ring}).
The velocity of each particle is determined in the beamline upstream of the ring.
The revolution time is obtained by dividing the measured TOF by the number of accomplished turns inside the ring.
The number of turns is then deduced from the revolution time measured 
by using the detector installed inside the ring
and the time interval between injection and extraction.
Relative precision of the TOF and velocity of $10^{-6}$ and $10^{-4}$, respectively,
are required for mass measurements with a relative precision of $10^{-6}$~\cite{ozawa}.
The confirmation of particle circulation, measurements of revolution times inside the ring, 
and simultaneous measurements of TOF and velocities are essential for successful mass measurements.

Conventional plastic scintillators 
were used as a stop detector for the TOF measurements in the ring and as a start detector for the velocity measurements.
We have developed two types of TOF detectors for use as a start detector for the TOF measurement in the ring: 
a stop detector for the velocity measurement,
and a diagnostics detector to confirm the circulation of particles~\cite{nagae, abe, suzuki}.
The required specifications are as follows:
(i) high time resolution, which is less than $50$~ps (standard deviation) for the velocity measurement,
less than $700$~ps for the TOF measurement, and less than $100$~ps for the circulation confirmation, 
as the flight times for the velocity measurements and for the TOF measurements are approximately $500$~ns and $0.7$~ms, respectively.
The typical revolution time is approximately $350$--$400$~ns.
(ii) A small amount of material to minimize the change and spread in particle velocity 
as a result of passing through the detector.
(iii) Preserving the charge state of the nuclei passing through the TOF detector to avoid particle loss.
(iv) A sufficiently large sensitive area to allow for a large beam size described later in this section.

We refer to one of the detectors as BE-MCP and to the other as E-MCP.
The working principle of BE-MCP is based on the perpendicular superposition of electric and magnetic fields
to achieve an isochronous transport of secondary electrons to a microchannel plate (MCP).
This type of detector was developed and successfully used in mass measurements at the Experimental Storage Ring (ESR)~\cite{feu, kno} 
and the experimental cooler storage ring (CSRe)~\cite{zhan, yhzhan} 
with a high time resolution of several tens of pico-seconds.
The BE-MCP detector developed for the Rare-RI Ring was tested by using an alpha source and a heavy ion beam.
We achieved a time resolution of approximately 40~ps and a detection efficiency of 95\%~\cite{suzuki}.
In the E-MCP, secondary electrons are transported to the MCP by using only electrostatic fields.
The same type of detector was developed for experiments with low-energy beams of 1--40~MeV/u,
such as studies of super heavy elements with the gas-filled recoil ion separator (GARIS) at RIKEN~\cite{morimoto, ishizawa},
elastic recoil detection experiments at the University of Jyv\"{a}skyl\"{a}~\cite{lait},
and fission fragment studies at GSI~\cite{kosev} and LANL~\cite{arnold}.

Both BE-MCP and E-MCP can be used as detectors at the ring entrance.
They provide a signal that is simultaneously used as the start signal for the TOF measurement in the ring and
as the stop signal for the velocity measurement.
Based on beam optics simulations, the beam diameter at the entrance of the ring is estimated 
to be approximately 30~mm corresponding to $\pm 3$ standard deviations of the Gaussian distribution ($\sigma$).
The E-MCP detector is better suited for confirming the particle circulation
inside the ring
because the magnetic field distribution of the ring is not changed between before and after removing the E-MCP detector 
from the beam path.
Furthermore, the E-MCP detector has advantages of easy operation and low construction cost as compared to the BE-MCP detector.
The injected particles circulate in the ring with betatron oscillations characterized
by the horizontal and vertical betatron functions, $\beta_x$ and $\beta_y$.
The $\beta$ functions at the E-MCP position 
are designed to be $\beta_x = 7.58$~m and $\beta_y = 10.52$~m in the horizontal and vertical directions, respectively.
Considering the momentum acceptance of the ring of $\pm 0.5$\% and
a dispersion of $70.5$~mm/\%, 
the beam diameter at the E-MCP detector position is estimated to be approximately $100$~mm ($\pm 3 \sigma$) and $50$~mm ($\pm 3 \sigma$) for the horizontal and vertical planes, respectively.

The E-MCP detector is intended to measure the TOF and velocity, as well as confirm the particle circulation
and measure the revolution time with high precision.
The revolution time is essential for determining the number of turns later used for mass determination.
For successful mass spectrometry at the Rare-RI Ring,
it is required to improve the time resolution and the detection efficiency 
as compared to the corresponding parameters of the existing detectors used in low-energy experiments.

To our knowledge, this study is the first attempt 
to apply an E-MCP type detector to a high-energy beam of a few hundred MeV/u.
This paper describes the design of the detector and its performance with a high-energy beam.
Furthermore, it provides the first results obtained for confirming particle circulation in the Rare-RI Ring.

\section{Description of the E-MCP detector}
\begin{figure}[htb]
\centering
\includegraphics[width=8cm]{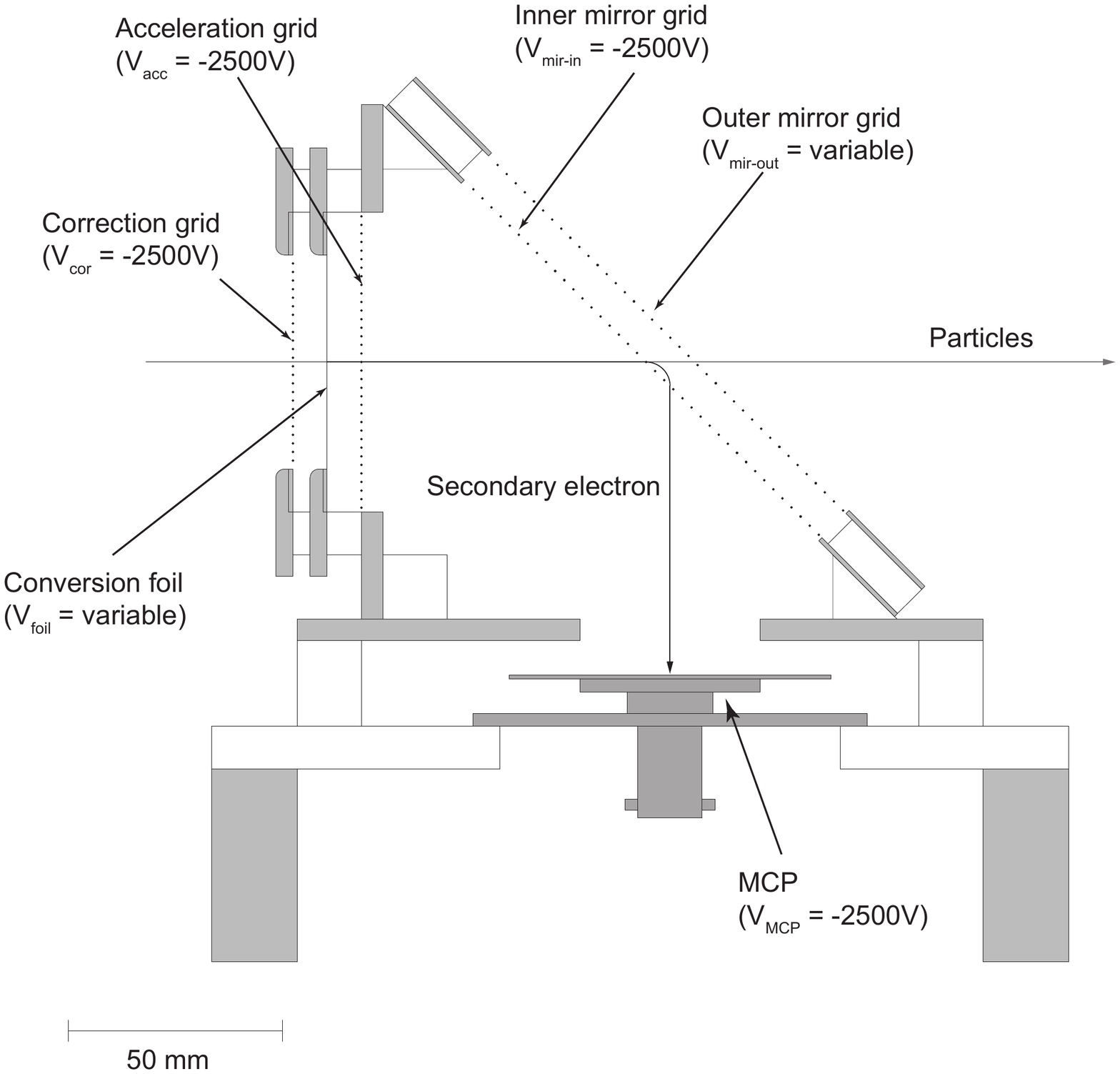}
\caption{Schematic view of the E-MCP detector. The white quadrilaterals and triangles are the Delrin insulators.}
\label{fig:E-MCP}
\end{figure}

A schematic view of the E-MCP detector and the potential arrangement are shown in Fig.~\ref{fig:E-MCP}.
The E-MCP detector consists of the conversion foil, the acceleration grid, the mirror grid, the MCP, and
the Delrin insulators.
Although Delrin has high outgassing rate, 
a vacuum pressure at the detector position in the ring was about $3 \times 10^{-5}$~Pa without any detectable effect of outgassing.
A vacuum pressure in the ring remained $2.5 \times 10^{-6}$ to $4.0 \times 10^{-5}$~Pa.
This pressure was sufficient to store the particles for approximately $2000$ turns, which corresponds to $0.7$~ms).
The E-MCP detector is not equipped with any magnetic shielding, 
because the magnetic fields at the detector position are as small as a few Gauss.

The secondary electrons are emitted from the conversion foil when a particle passes through the foil. 
Typical electron energies are a few eV~\cite{hassel}. 
They are accelerated by the acceleration electric potential applied between the conversion foil and the acceleration grid,
such that the initial energies and angular spreads become negligible.
As a result, the initial conditions have marginal effect on the time resolution and detection efficiency.
After the acceleration, the secondary electrons drift through the field-free region to the electrostatic mirror grid 
and are subsequently deflected by 90$^\circ$ toward the MCP by the electrostatic mirror field
applied between the inner and the outer mirror grids.

The used conversion foil was either a $60$-$\mu$g/cm$^{2}$-thick carbon foil developed at RIKEN~\cite{hasebe}
or an aluminum-coated mylar foil ($0.1$-$\mu$m-thick aluminum coated on both sides with $2$~$\mu$m mylar).
The yield of secondary electrons is calculated to be reduced 
with increasing the incident beam energy by using a semi-empirical theory of electron emission as described in Section \ref{eff}.
An enhancement of the yield is essential in the high-energy beam experiment.
This enhancement of the yield is decisive for the improvements of the time resolution and the detection efficiency.
To obtain a sufficient yield of secondary electrons for the high-energy beam, 
the thick foils were used in the E-MCP.

The carbon foil became deformed after its production,
while the aluminum-coated mylar foil remained homogeneously flat.
Based on the beam sizes estimated by ion-optical calculations,
the size of both foils was fixed to $100$~mm in the horizontal direction and $50$~mm in the vertical direction. 
The relative velocity change for $^{78}$Ge at $E = 170$~MeV/u is estimated
to be $9.18 \times 10^{-6}$ in the carbon foil 
and $4.55 \times 10^{-5}$ in the aluminum-coated mylar foil per passage using the ATIMA code~\cite{atima}. 
The relative velocity spreads (standard deviation) are $2.44 \times 10^{-6}$ and $4.50 \times 10^{-6}$ 
for the carbon and the aluminum-coated mylar foils, respectively.
These values ensure a relative velocity precision in the order of $10^{-4}$.
The percentage of particles that preserve their charge state 
upon passage through the aluminum-coated mylar foil was estimated to be $99.2$\% 
for the fully striped $^{78}$Ge ion at $E = 170$~MeV/u by using the GLOBAL code~\cite{global}.

The acceleration grid is made of gold-plated tungsten wires with a diameter of $40$~$\mu$m, 
set at 1-mm pitch to create a homogeneous acceleration electric field, 
which is essential for a good time resolution.
The grid is placed at a distance of $8$~mm from the conversion foil.
The geometrical transparency of the acceleration grid is approximately $96$\%. 
The mirror is formed by the inner and the outer mirror grids installed $8$~mm apart from each other
and tilted by $45^\circ$ with respect to the conversion foil.
These grids are also made of the same gold-plated tungsten wires but with a $3$-mm pitch
to achieve a larger geometrical transparency of approximately $99$\%.
Therefore, the total geometrical transparency for the E-MCP detector is $94$\%.
A large electric field may cause tension on the conversion foil,
which may lead to deformation or even damage.
In order to prevent this, a correction grid made of the same wires with a $1$-mm pitch 
is set opposite to the acceleration grid and held under the same potential as the acceleration grid.
This correction grid is only used with the carbon conversion foil,
which is susceptible to such additional deformation under a strong acceleration field (higher than $200$~V/mm).
In this case, the total geometrical transparency of the E-MCP detector is reduced to approximately $90$\%.

In a previous study~\cite{abe}, we employed a rectangular chevron-type MCP (95~mm $\times$ 42~mm) 
to match the beam size inside the ring.
The obtained time resolution of 130~ps and the detection efficiency of 72\% for the $^{84}$Kr beam of 170~MeV/u were insufficient.
In this work, another chevron-type MCP (Hamamatsu Photonics F9892-14) was applied for the detection of secondary electrons.
Its use inside the ring reduced the detection efficiency, 
since its sensitive area ($42$~mm in diameter) is small as compared with the estimated beam size.
The timing performance of the MCP was, however, sufficient for measuring the revolution time, the TOF in the ring, 
and the velocity of the nuclei.

Based on the investigations on the detection efficiency by Boccaccio et al.~\cite{bocca},
the acceleration field ($E_\mathrm{acc}$) and the mirror field ($E_\mathrm{mir}$) were chosen 
to satisfy the relation $2E_\mathrm{acc} =  E_\mathrm{mir}$.
The potentials of the conversion foil and the outer mirror grid were optimized accordingly.
The potentials of the acceleration grid and the inner mirror grid were set to the same potential 
as the MCP front of $-2500$~V.

\section{Performance of the E-MCP detector}
\subsection{Test experiment}

\begin{figure}[htb]
\centering
\includegraphics[width=8.0cm]{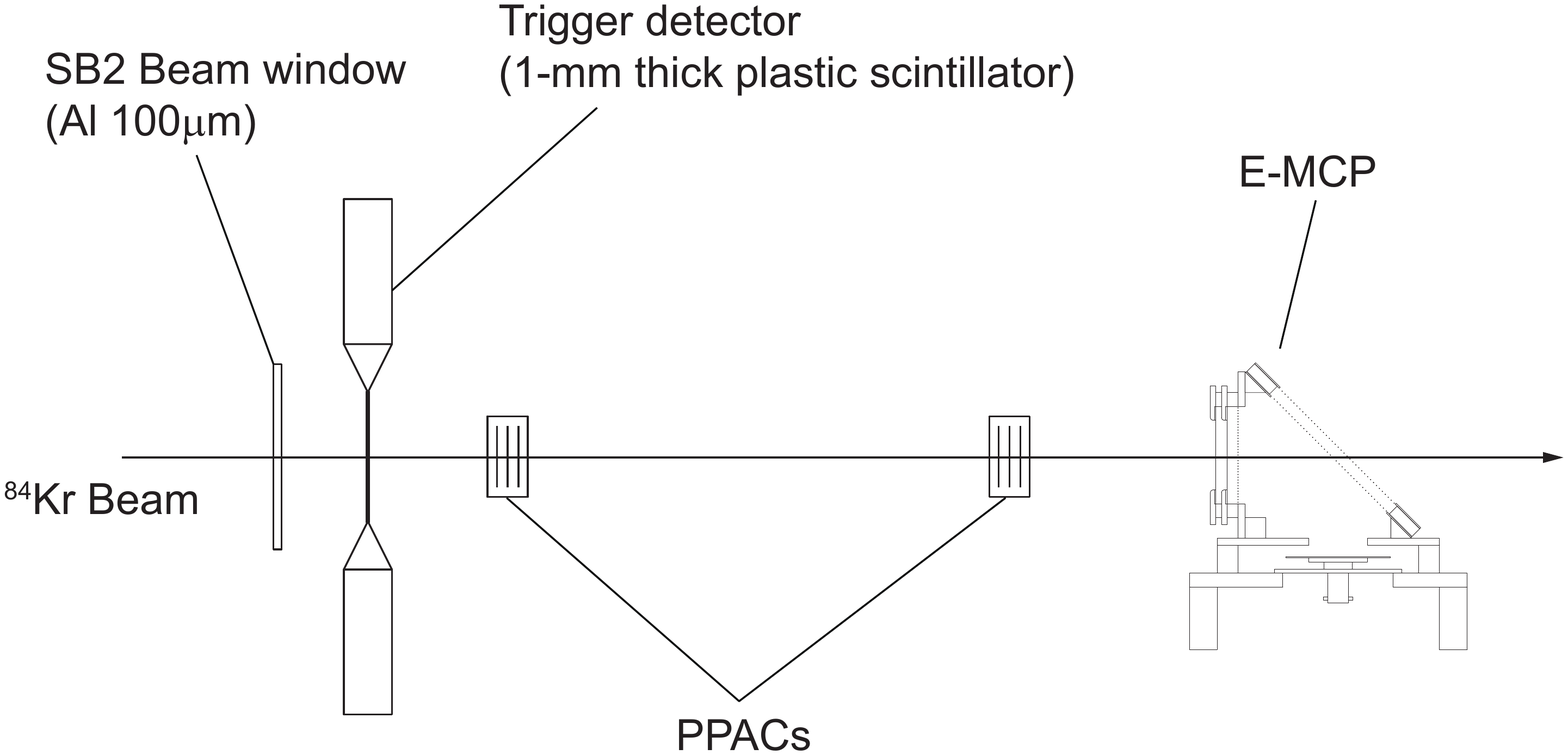}
\caption{Schematic view of the setup using heavy-ion beams at HIMAC.}
\label{fig:setup}
\end{figure}

Experiments to investigate the performance of the E-MCP detector were performed 
at the SB2 course~\cite{kanazawa} of the heavy-ion synchrotron facility, 
Heavy Ion Medical Accelerator in Chiba (HIMAC)~\cite{hirao}, National Institute of Radiological Sciences. 
Figure~\ref{fig:setup} shows a typical experimental setup.
A heavy-ion beam of $^{84}$Kr ($E = 200$~MeV/u) 
was transported to the final focus of the SB2 course, 
with a typical intensity of $2 \times 10^{3}$ particles per spill.
The $1$-mm-thick plastic scintillator was located behind a $100$-$\mu$m-thick aluminum vacuum window in SB2.
Both ends of the scintillator were optically connected to photomultiplier tubes.
It was used as the trigger detector.
The E-MCP detector was placed approximately $1$~m downstream of the trigger detector.
Two parallel plate avalanche counters (PPAC)~\cite{kumagai} were used to measure the position distribution of the beam 
and to estimate the beam profile at the conversion foil using a ray-trace technique. 
The typical beam diameter on the conversion foil was $18$~mm ($\pm 3 \sigma$).
The energy of the $^{84}$Kr beam at the E-MCP detector was calculated 
to be approximately $E \sim$ $170$~MeV/u using the ATIMA code~\cite{atima} .

Standard electronics was used in this test experiment. 
The signals from the trigger detector and the MCP were processed
by using a constant fraction discriminator module (Ortec935).
The time interval between trigger detector signal and the MCP signal was converted to 
an analog output pulse 
by using a time-to-amplitude converter (Ortec567) with full-range of 50-ns.
The analog output pulse was converted into a digital form by an analog-to-digital converter.

We investigated the time resolution and the detection efficiency
in dependence from the applied acceleration potential for each conversion foil.

\subsection{The time resolution}
\begin{figure}[tb]
\centering
\includegraphics[width=8.0cm]{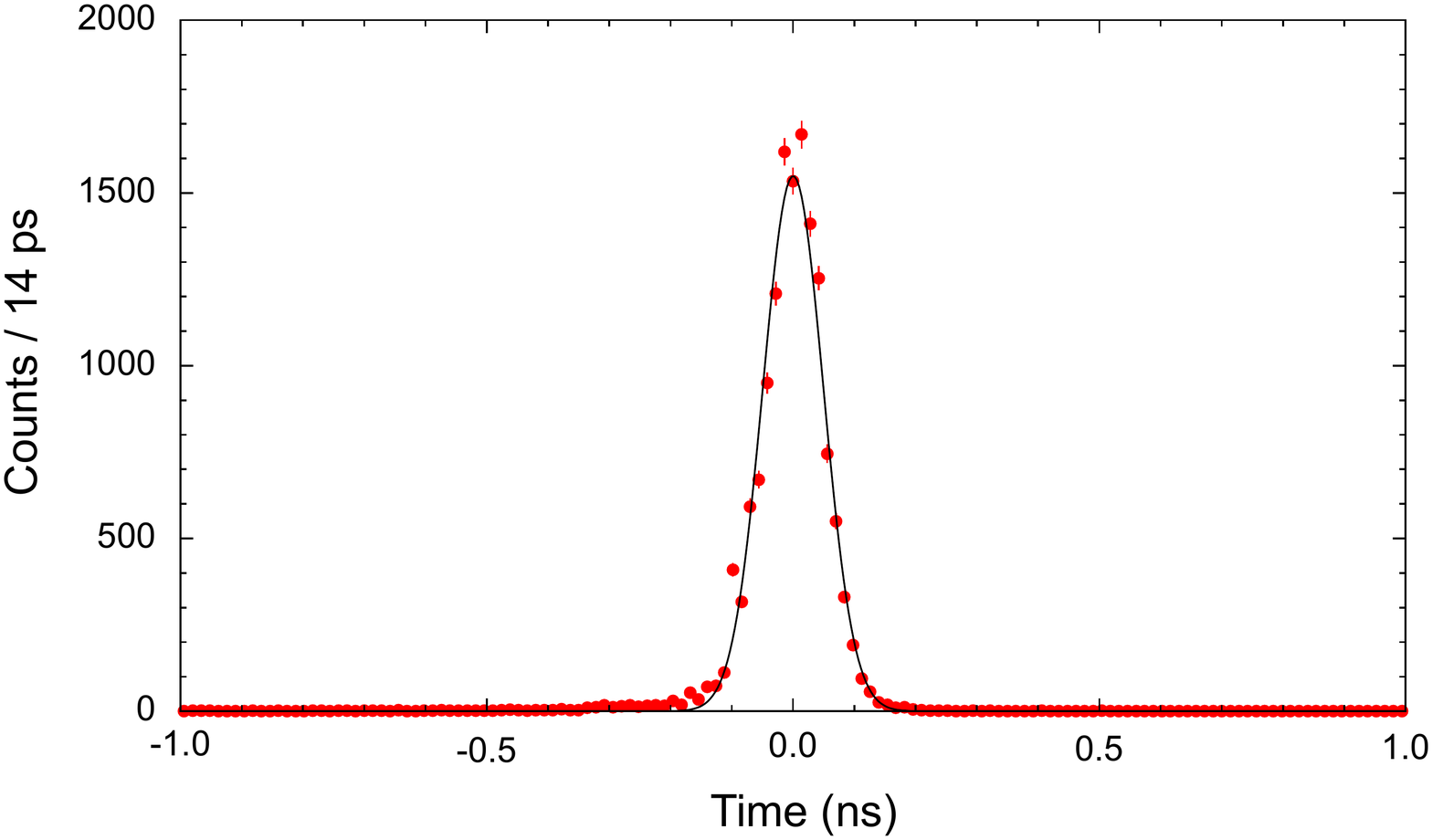}
\caption{Time spectrum of the E-MCP detector obtained at $U_\mathrm{acc} = 1500$~V. 
The solid line shows the Gaussian function providing the best fit for the collected data.
The width of the peak is $\sigma_\mathrm{peak} = 49.7(3)$~ps.}
\label{fig:e-mcp-spe}
\end{figure}

To investigate the time resolution of the E-MCP detector, 
we measured the TOF between the trigger detector and the E-MCP detector.
A typical TOF spectrum, using the aluminum-coated mylar as the conversion foil, is shown in Fig.~\ref{fig:e-mcp-spe}. 
The width of the peak ($\sigma_\mathrm{peak}$) was determined by fitting the TOF spectrum with a Gaussian function.
This value includes the intrinsic time resolution of the E-MCP detector ($\sigma_\mathrm{E-MCP}$),
the intrinsic time resolution of the trigger detector ($\sigma_\mathrm{trigger}$), 
and the timing jitter from the electronics ($\sigma_\mathrm{jitter}$).
The intrinsic time resolution of the trigger detector and the timing jitter
were measured to be $\sigma_\mathrm{trigger} = 18(1)$~ps and $\sigma_\mathrm{jitter} = 15(1)$~ps, respectively. 
From these values, we were able to calculate the intrinsic time resolution of the E-MCP detector via: 
$\sigma_\mathrm{E-MCP} = \sqrt{ \sigma_\mathrm{peak}^{2} - \sigma_\mathrm{trigger}^{2} - \sigma_\mathrm{jitter}^{2} }$.

\begin{figure}[tb]
\centering
\includegraphics[width=8.0cm]{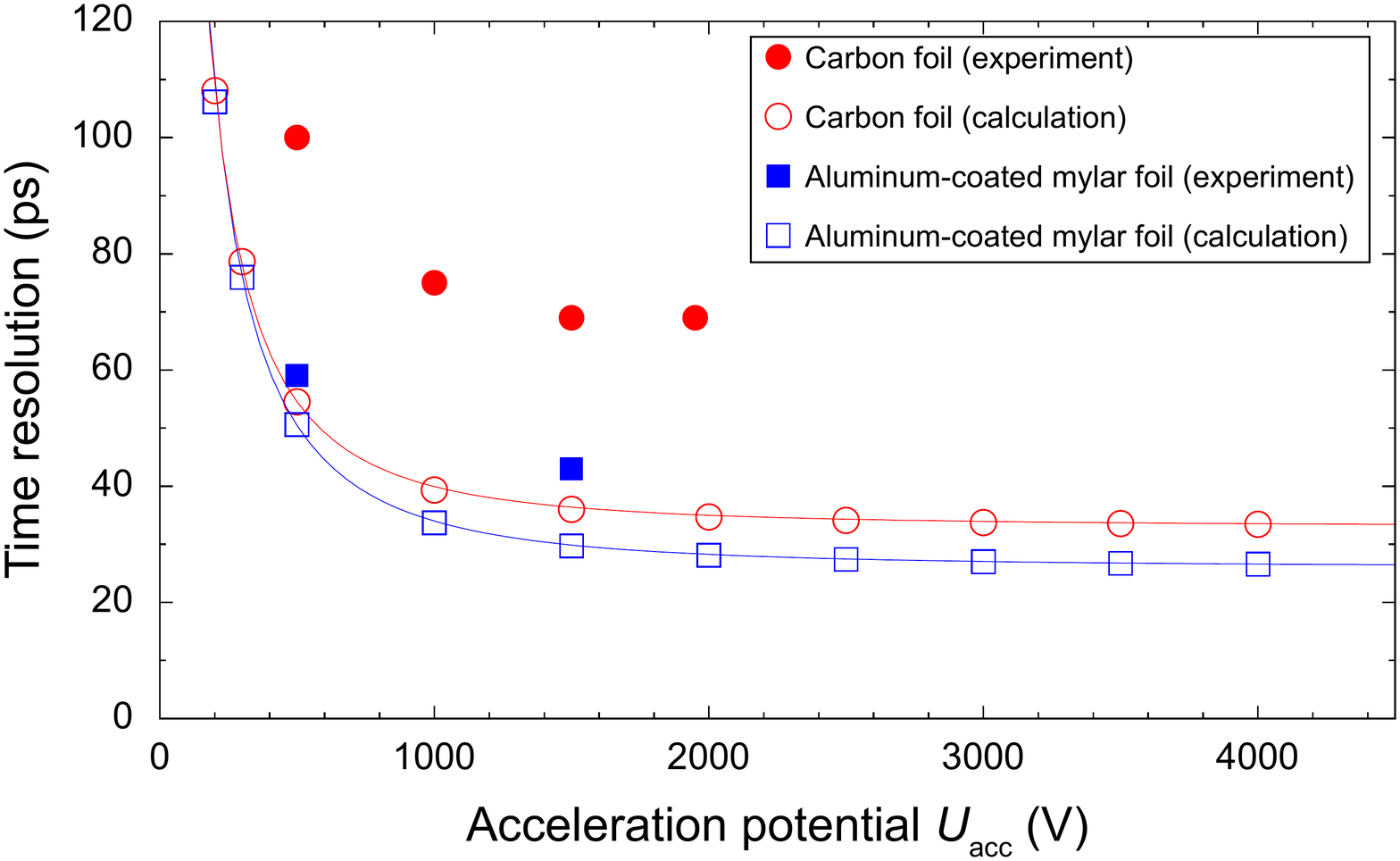}
\caption{The intrinsic time resolution of the E-MCP detector as a function of the acceleration potential.
The filled red circles and filled blue squares represent the experimental time resolutions when using the carbon 
and the aluminum-coated mylar foils, respectively.
All error bars are smaller than the symbol size and thus not visible in this figure.
The open red circles and blue squares represent the calculated time resolutions for the carbon 
and the aluminum-coated mylar foils, respectively.
The solid lines connecting the calculated results are provided to guide the eye.}
\label{fig:resolution}
\end{figure}

Figure \ref{fig:resolution} shows the variation of experimental intrinsic time resolution of the E-MCP detector
with the acceleration potential ($U_\mathrm{acc} = V_\mathrm{acc} - V_\mathrm{foil}$) with either the carbon foil (red circles) 
or the aluminum-coated mylar foil (blue squares).
The intrinsic time resolution obtained with the aluminum-coated mylar foil was better than that obtained with the carbon foil. 
We assume that this discrepancy may arise from the developed static deformation of the carbon foil,
 which causes variations of the acceleration potential and of the initial flight direction of secondary electrons.

The time resolution of the TOF detector using secondary electrons emitted from a foil system 
depends on the acceleration potential~\cite{zhan, lait, mizota}.
The value of the experimental intrinsic time resolution decreased with an increase in the acceleration potential for both foils. 
Figure \ref{fig:resolution} shows the calculated intrinsic time resolution combined with a 3D simulation performed 
by using the SIMION package~\cite{simion}
and transit time spreads in the MCP, estimated to be $33$~ps and $26$~ps~\cite{fra} 
for the carbon and the aluminum-coated mylar foils, respectively.
The transit-time spread in the MCP depends on the number of incident secondary electrons.
As described in Section \ref{eff}, the yield of the secondary electrons emitted from the foil
was estimated by using Eq.~\ref{eq:gamma} and depends on the energy and the charge of the particle and the foil thickness.
The difference between the calculated intrinsic time resolution with the carbon foil and that with the aluminum-coated mylar foil
is essentially due to the difference in the transit time spreads in the MCP.
In this simulation, the initial position distribution was set to $\sigma=3$~mm to reproduce the experimental conditions.
The initial energy was set to $E = 2$~eV (FWHM = $1$~eV)~\cite{hassel, hass}, 
and the initial angle was set to be perpendicular to the conversion foil.
An isotropic distribution of the electrons was assumed.
The grids were substituted in this simulation with a $0$-mm-thick foil 
to obtain an extremely homogeneous electric field.
Thus, the calculated time resolutions are considered ideal values.
We note, that the calculated time resolutions are better than the experimental ones, 
possibly owing to the modelization of the grid.
Although the experimental time resolutions do not coincide with the calculated values,
the intrinsic time resolutions of $\sigma_\mathrm{E-MCP} = 69(1)$~ps and $\sigma_\mathrm{E-MCP} = 43(1)$~ps were
experimentally achieved at $U_\mathrm{acc} = 1500$~V
for the carbon and the aluminum-coated mylar foils, respectively.
The obtained time resolutions are comparable 
or better than those of the detectors employed in the low-energy beam experiments.
These obtained resolutions of the E-MCP detector are sufficient for evaluating the revolution times inside the Rare-RI Ring,
the TOF in the ring, and the velocity determination upstream the ring.

\subsection{The detection efficiency}
\label{eff}
The detection efficiency of the E-MCP detector is defined by the ratio of the number of incident particles at the conversion foil 
and the number of signals detected by the MCP.
The number of incident particles at the conversion foil was obtained from the estimated beam profile.
The number of detected signals depends on the secondary electron yield, 
the detection efficiency of the MCP, and the transport efficiency for the secondary electrons~\cite{suzuki}.

The secondary electron yield per incoming particle has been well studied~\cite{rot1995,jung}.
The estimation is based on a semi-empirical theory of electron emission:
the electron yield in the forward direction with respect to the beam is given as;
\begin{equation}
\label{eq:gamma}
\gamma = \Lambda \frac{dE}{dx}\left[ 1-\beta_\mathrm{s} exp\left(-\frac{d}{\lambda_\mathrm{s}}\right) -\beta_{\delta} exp\left(-\frac{d}{\lambda_{\delta}}\right) \right],
\end{equation}
where $\Lambda$ is a constant that depends mainly on the foil material,
$dE/dx$ the energy loss per unit path length in units of keV/($\mu$g/cm$^{2}$),
$\beta_\mathrm{s}$ and $\beta_{\delta} = 1 - \beta_\mathrm{s}$ partition factors of ion energy loss for secondary electrons
and $\delta$ electrons, respectively,
$d$ denotes the foil thickness,
$\lambda_\mathrm{s}$ and $\lambda_{\delta}$ are diffusion lengths for secondary electrons and $\delta$ electrons, respectively.

The yield of secondary electrons emitted from the carbon foil was calculated to be $10$ electrons,
adopting the beam energy of $E = 170$~MeV/u, $\Lambda$ for carbon $\Lambda_\mathrm{C} = 7$~$\mu$g/cm$^{2}$/keV~\cite{rot1990}, $\beta_{\delta} = 0.75$, 
$\lambda_\mathrm{s} = 19$~\AA, and $\lambda_{\delta} = 2.05 \times 10^{5}$~\AA.
These values, except for $\Lambda_\mathrm{C}$, were taken from Ref.~\cite{jung}.
$\lambda_{\delta}$ for carbon was calculated using $\lambda_{\delta} = 390E^{1.22}$~\cite{jung}.

In the case of the aluminum-coated mylar foil, the secondary electron yield was calculated to be $20$ electrons
assuming $E = 170$~MeV/u, $\Lambda$ for aluminum $\Lambda_\mathrm{Al} = 13$~$\mu$g/cm$^{2}$/keV~\cite{clo}, 
$\beta_{\delta} = 0.75$~\cite{jung}, 
$\lambda_\mathrm{s} = 16.4$~\AA~\cite{bro}, and $\lambda_{\delta} = 1.51 \times 10^{4}$~\AA.
$\lambda_{\delta}$ was calculated by using a ratio of $\lambda_{\delta}/\lambda_\mathrm{s} \sim 5.4E$~\cite{rot1990}.
Note, that the ratio was obtained for a low-energy beam.
$\lambda_{\delta}$ depends on the energy of the $\delta$ electrons,
while the energy of the $\delta$ electrons depends on the energy of the incident beam.
Assuming that the energy of the $\delta$ electrons from aluminum is the same for those from carbon, 
$\lambda_{\delta}$ was calculated to be $\lambda_{\delta} = 1.84 \times 10^{5}$~\AA~using the same method described in Ref.~\cite{suzuki}.
As a result, the secondary electron yield was calculated to be $17$ electrons.
The difference in the secondary electron yields between the two calculated results is small
and does not affect the estimated detection efficiency of the detector.

The detection efficiency of the MCP depends on the energy of the secondary electrons 
and reaches a plateau at 75\% for electron impact energy above 500~eV~\cite{bowm}.
In the E-MCP detector, the electron impact energy was tuned to be larger than 500~eV.
By considering an assumed transport efficiency of 94\%, obtained with SIMION, 
and an open area ratio of the MCP device of 60\%,
the efficiency of the E-MCP detector in the detection of more than one electron was estimated to be approximately 100\% for both foils.
The same equations as those described in Ref.~\cite{suzuki} were used for this estimation.

\begin{figure}[tb]
\centering
\includegraphics[width=8.0cm]{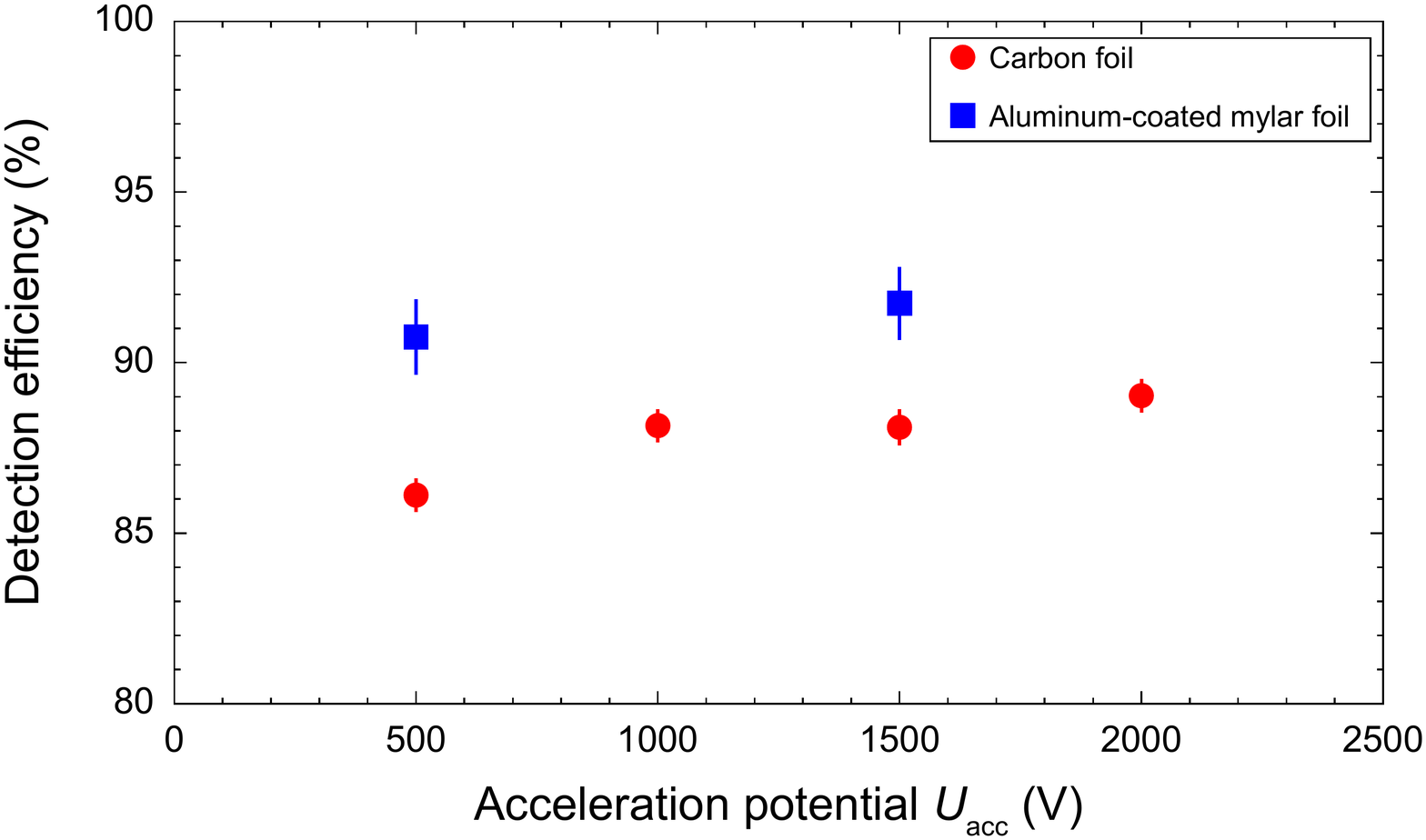}
\caption{The detection efficiency of the E-MCP detector as a function of the acceleration potential.
The red circles and blue squares represent experimental detection efficiencies
 when using the carbon foil and the aluminum-coated mylar foil, respectively.}
\label{fig:eff}
\end{figure}

The experimental detection efficiencies are shown in Fig.~\ref{fig:eff} as a function of the acceleration potential.
In contrast to the results obtained for the time resolution, 
we did not find any dependence of the detection efficiencies on the acceleration potential.
The efficiencies remain almost constant at approximately 90\%.
The detection efficiencies achieved were 88(1)\%  and 92(1)\%
for the applied $U_\mathrm{acc}$ = 1500~V for the carbon and the aluminum-coated mylar foils, respectively.
The discrepancy between the experimental detection efficiencies and the calculated values of approximately 100\%
is believed to be due to small amplitudes of the signals created in cases with a small number of secondary electrons.
The small-amplitude signals were not accepted by the electronics
and the obtained detection efficiencies were reduced compared with the calculation.
Although the experimental detection efficiencies do not reach 100\%, 
their magnitudes are sufficient for use in Rare-RI Ring experiments.

\section{Application in mass measurements at Rare-RI Ring}
In 2016, the first demonstration of a mass measurement was performed 
using exotic nuclei with well-known masses around $^{78}$Ge~\cite{storinagae}.
Exotic nuclei were obtained from the in-flight fission of a $^{238}$U relativistic primary beam 
with an energy of 345~MeV/u on a 10-mm-thick $^{9}$Be target. 
Particles of $^{79}$As, $^{78}$Ge, $^{77}$Ga, $^{76}$Zn, and $^{75}$Cu 
were transported to the ring, and then injected into the ring one by one 
employing the individual particle injection method with a fast-response kicker system~\cite{miura}.
The particle circulation was confirmed by using the E-MCP detector.
Measurements of the revolution times were also performed.
These revolution times were then used to obtain the corresponding numbers of accomplished revolutions 
in mass determination.
The carbon foil was selected as the conversion foil,
since the energy loss in the carbon foil is smaller as compared to that of the aluminum-coated mylar foil.
After the circulation confirmation, the E-MCP detector was removed from the ring aperture.
Then, a precise tuning of the magnetic field of the ring was performed 
to achieve the isochronous condition for the reference particle of $^{78}$Ge.
The TOF in the ring and the velocity were simultaneously measured for mass determination 
by using another E-MCP detector equipped 
with the carbon foil placed at the entrance of the ring.

\begin{figure}[tb]
\centering
\includegraphics[width=8.0cm]{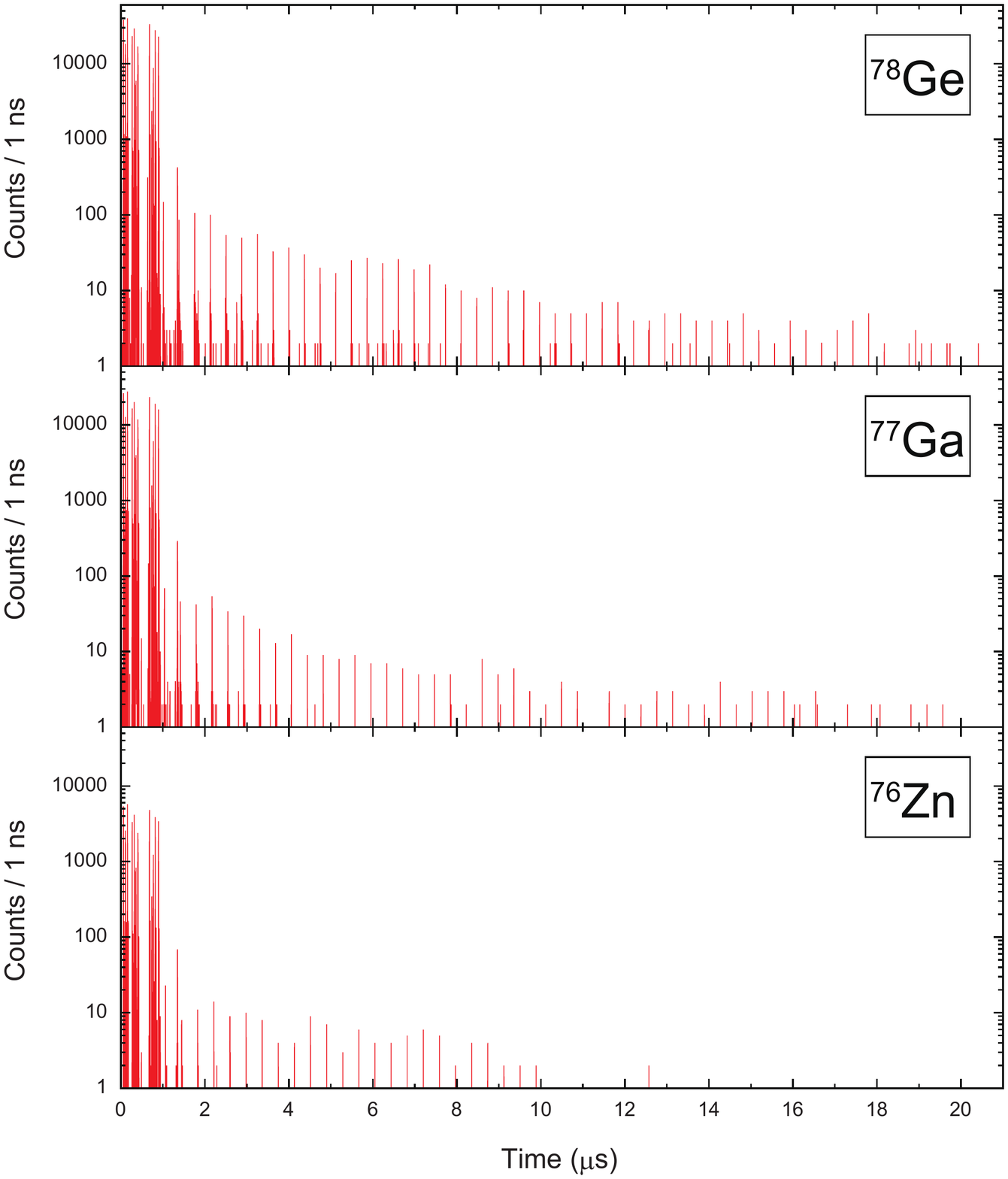}
\caption{Time signals of $^{78}$Ge, $^{77}$Ga, and $^{76}$Zn.
Spurious signals generated by the noise from the kicker magnet are observed at $0 \sim 1$~$\mu$s duration.}
\label{fig:storage}
\end{figure}

The E-MCP detector used for the circulation confirmation was equipped with the sufficiently large carbon foil
to cover the required sensitive area.
As described before, since the sensitive area of the MCP is only $42$~mm in diameter,
the E-MCP detector can only detect particles near the central orbit.
The potentials of the carbon foil and of the outer mirror grid were set to $-3000$~V and $-3500$~V, respectively,
because the time resolution at $U_\mathrm{acc} = 500$~V was sufficient to measure the revolution times.
The correction grid was not used in this measurement
because the $E_\mathrm{acc}$ value was as small as $62.5$~V/mm.
Each stored particle passed through the foil of the E-MCP detector at each revolution in the Rare-RI Ring.
The MCP signals were discriminated by using the constant fraction discriminator module (Ortec935) in the same way as in the test experiment. 
To digitize the signal from the E-MCP detector, 
a multi-hit time-to-digital converter (Acqiris TC890) was used.
Time signals for the stored $^{78}$Ge, $^{77}$Ge, and $^{76}$Zn particles are shown in Fig.~\ref{fig:storage}.
Periodic time signals were clearly observed in the spectra.
Since the geometrical transparency of the E-MCP detector is only 94\%,
the particles might collide with the grids of the E-MCP detector and lose energy.
Therefore, the number of turns was limited to a few tens.
This result indicates the successfully achieved circulation confirmation.

\begin{figure}[tb]
\centering
\includegraphics[width=8.0cm]{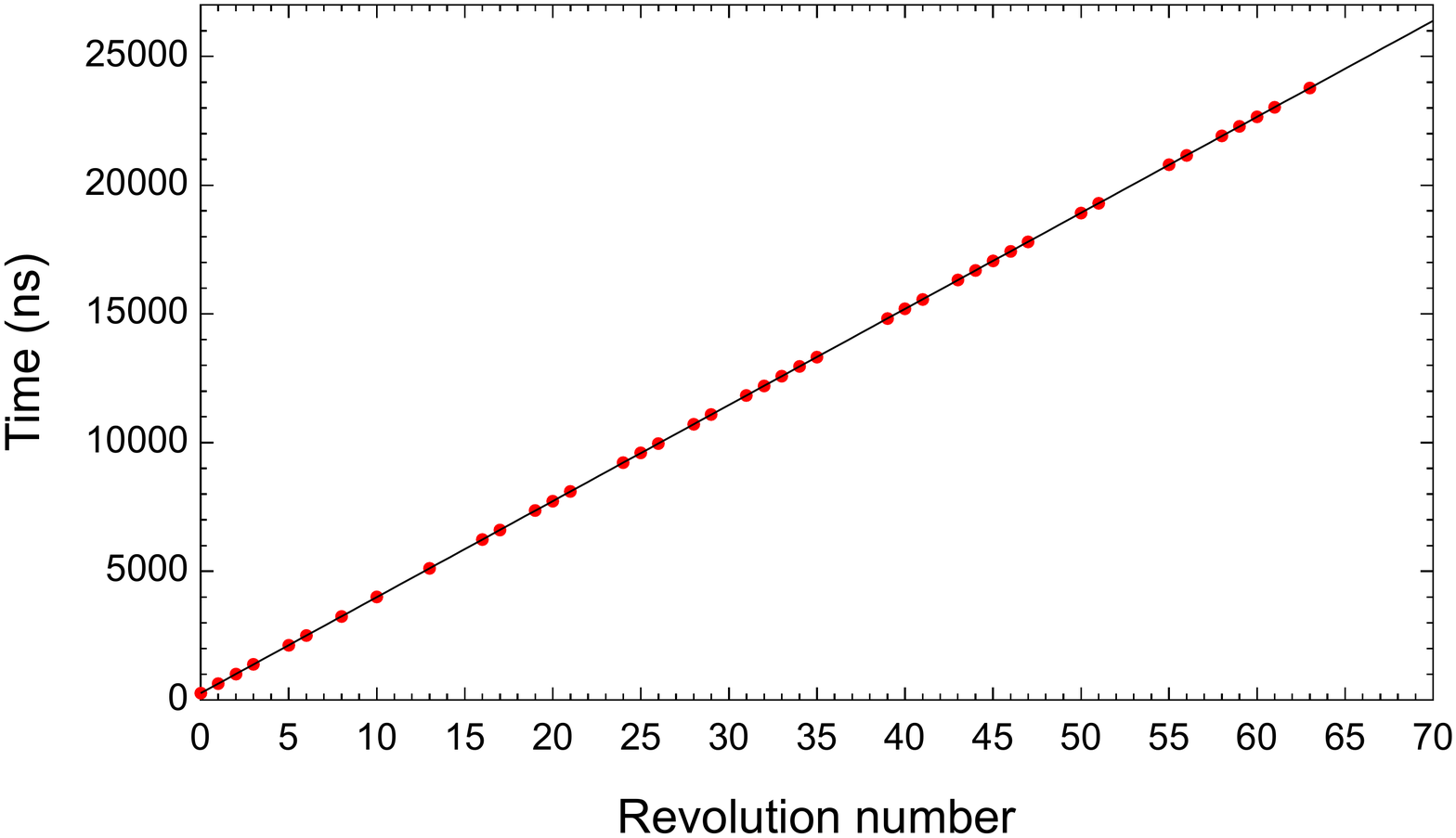}
\caption{Passage time of one $^{78}$Ge particle as a function of the revolution number.
The line is a straight-line fit through the data points.}
\label{fig:78ge-194}
\end{figure}

\begin{figure}[tb]
\centering
\includegraphics[width=8.0cm]{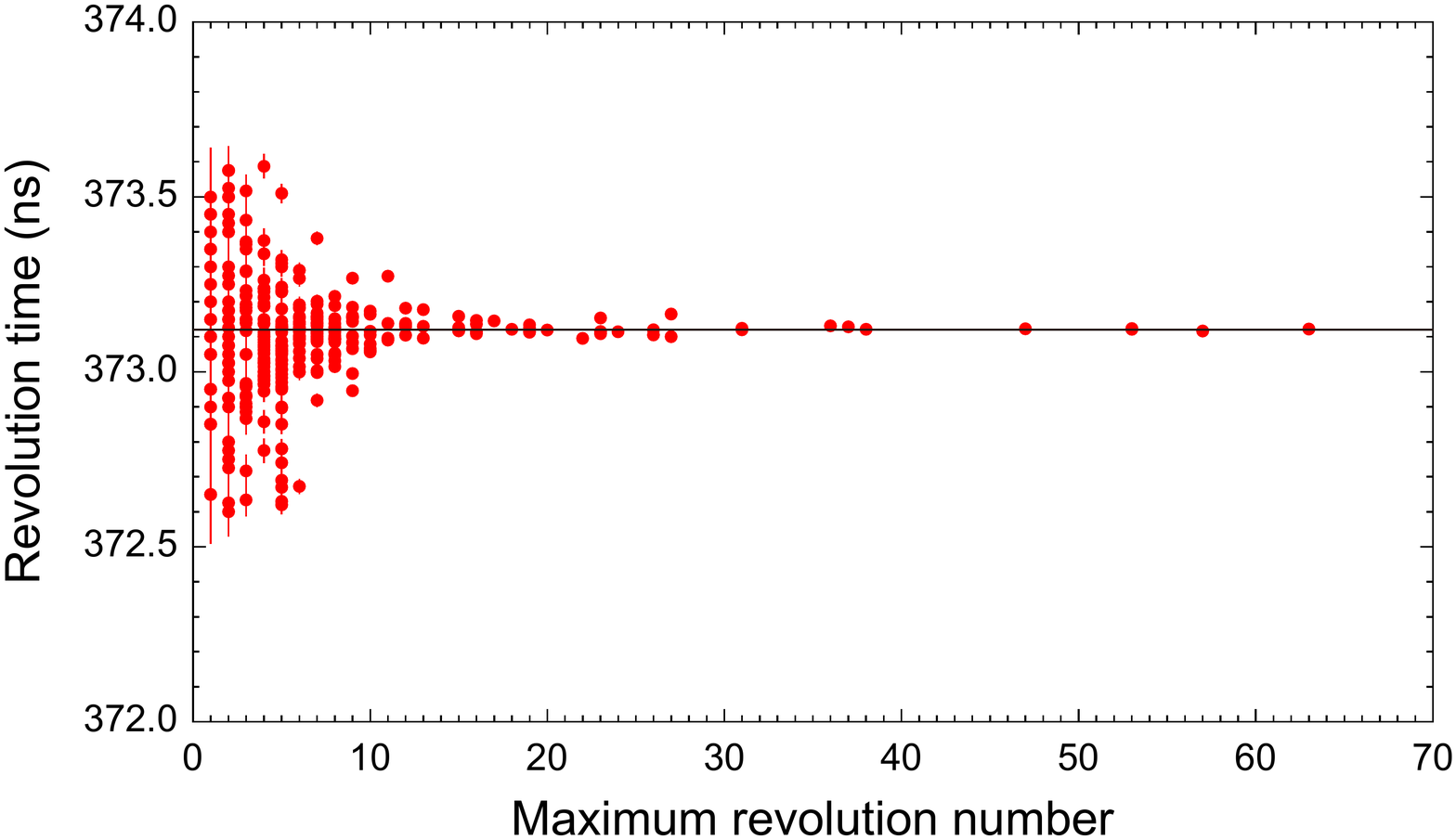}
\caption{Measured revolution times for $^{78}$Ge. The line indicates the mean revolution time.}
\label{fig:revo}
\end{figure}

The revolution times were obtained from an event-by-event analysis.
For example, the passing time for one event of $^{78}$Ge is shown in Fig.~\ref{fig:78ge-194} 
as a function of the revolution number.
The revolution time slightly changed turn-by-turn due to the energy loss in the foil.
In order to obtained the revolution time, we fitted the data by using first, second and third-order polynomial functions.
The residuals were almost the same.
Furthermore, no dependencies in the residuals on the revolution number were observed.
Thus, the revolution time was obtained by fitting the passing times 
as a function of the revolution number with a first-order polynomial.
The last timing signal was used to deduce the maximum revolution number for each stored particle.
The deduced revolution times of each event for $^{78}$Ge are shown in Fig.~\ref{fig:revo} 
as a function of the maximum revolution number.
The revolution times for the events with small maximum revolution numbers show a larger scatter.
This result suggests that the particles with a larger emittance circulate for fewer turns.
With increasing the maximum revolution number, the revolution time converges to a constant value.
By considering a weighted average of these revolution times, we obtained a mean revolution time value of 373.121(2)~ns for $^{78}$Ge.
The revolution times of $^{77}$Ga and $^{76}$Zn obtained by adopting the same analysis 
were found to be 378.192(3)~ns and 383.615(5)~ns, respectively.
The obtained relative precisions were on the order of $10^{-5}$ to $10^{-4}$.
Taking into account the number of turns in the mass measurement of approximately $2000$,
the achieved precision is sufficient to obtain the exact number of turns 
for each particle in the mass measurement at Rare-RI Ring.

\section{Conclusion}
In this paper, we presented the design of a TOF detector named E-MCP
and evaluated its performance with the high-energy heavy-ion beams.
It was employed to confirm the particle circulation at the Rare-RI Ring in RIKEN.
The dependence of the time resolution and the detection efficiency 
on the acceleration potential and on the choice of conversion foil 
were systematically evaluated by using $^{84}$Kr at an energy of 170~MeV/u at HIMAC.
We achieved the time resolutions (standard deviation) of  
$\sigma_\mathrm{E-MCP}$ = 69(1)~ps and $\sigma_\mathrm{E-MCP}$ = 43(1)~ps 
when using the carbon and the aluminum-coated mylar foils, respectively. 
The detection efficiencies were found to be 88(1)\% and 92(1)\% 
for the carbon and the aluminum-coated mylar foils, respectively. 
These results indicate the E-MCP type detector has the sufficient performance 
not only for the low-energy beam experiments but also for the high-energy beam experiments.
Although its performance was inferior to simulated and calculated results,
the obtained performance was sufficient for use in the mass-measurement experiments at the Rare-RI Ring.
The experiments with the E-MCP detector with the carbon foil
were able to successfully confirm the particle circulation
for up to 60 turns 
and to deduce corresponding revolution times with high precision.
The successful revolution-time measurements lead to obtain the exact number of turns 
and make the mass measurement accurate.

\section{Acknowledgements}
The authors thank the accelerator staff of NIRS-HIMAC and the staff of the RIKEN Ring Cyclotron 
for their support during the experiment. 
Performance tests were supported by Research Project with Heavy Ions at NIRS-HIMAC.
Commissioning experiments of the Rare-RI Ring were performed at the RI beam factory operated 
by RIKEN Nishina Center and CNS, University of Tokyo 
under Experimental Program MS-EXP15-04, MS-EXP15-12 and MS-EXP16-10. 
These were supported by the RIKEN Pioneering Project funding 
and JSPS KAKENHI Grant Numbers JP25105506, JP26287036, JP15H00830, and JP17H01123.
YAL acknowledges support from the European Research Council (ERC) under the European Union's Horizon 2020 research
and innovation programme [Grant Agreement Number 682841 ``ASTRUm''].

\bibliography{e-mcp}
\end{document}